\documentstyle{article}
\everymath{\rm }
\everydisplay{\rm }
\begin{document}
\begin{center}
{\LARGE Total intrinsic spin  \\ [.125in]
for plane gravity waves} \\ [.25in]
\large Donald E. Neville \footnote{\large Electronic address:
nev@vm.temple.edu }\\Department of Physics \\Temple University
\\Philadelphia 19122, Pa. \\ [.25in]
March 28, 1997 \\ [.25in]
\end{center}
%%%%%%Ashtekar Notations %%%%%%%%%%%
\newcommand{\E}[2]{\mbox{$\tilde{{\rm E}} ^{#1}_{#2}$}}
\newcommand{\A}[2]{\mbox{${\rm A}^{#1}_{#2}$}}
\newcommand{\Np}{\mbox{${\rm N}'$}}
\newcommand{\Etwo}{\mbox{$^{(2)}\!\tilde{\rm E} $}\ }
\newcommand{\Etld }{\mbox{$\tilde{\rm E}  $}\ }
\def \ut#1{\rlap{\lower1ex\hbox{$\sim$}}#1{}}
\newcommand{\phst}{\mbox{$\phi\!*$}}
\newcommand{\psist}{\mbox{$\psi\!*$}}
%%%%%%%Tex Newcommands%%%%%%%%%
\newcommand{\bea}{\begin{eqnarray}}
\newcommand{\eea}{\end{eqnarray}}
\newcommand{\be}{\begin{equation}}
\newcommand{\ee}{\end{equation}}
\newcommand{\nn}{\nonumber \\}
\newcommand{\rta}{\mbox{$\rightarrow$}}
\newcommand{\rla}{\mbox{$\leftrightarrow$}}
\newcommand{\eq}[1]{eq.~(\ref{eq:#1})}
\newcommand{\Eq}[1]{Eq.~(\ref{eq:#1})}
\newcommand{\eqs}[2]{eqs.~(\ref{eq:#1}) and (\ref{eq:#2})}
\large
\begin{center}
{\bf Abstract}
\end{center}
A quantity which measures total intrinsic spin along the z axis
is
constructed for planar gravity
(fields dependent on z and t only), in both the Ashtekar
complex connection formalism and in geometrodynamics.  The total
spin  is conserved but (surprisingly) is not a surface
term.  This constant of the motion  coincides with one of four
observables
previously discovered by Husain and Smolin.  Two more of those
observables can be interpreted physically as raising and lowering
operators for total spin.  \\[.125in]
PACS categories: 04.60, 04.30
\clearpage
\section{Introduction}
\label{int}

     This paper derives a constant of the motion which gives the
total spin angular momentum along the z axis, for planar
symmetric
gravitational waves moving in the directions $\pm$ z.  This
constant of the motion was derived as part of an ongoing
investigation \cite{I,II,III} of the properties of the complex
connection formalism proposed by Ashtekar \cite{Ash87}, and the
derivation is carried out within that formalism; however, the
final
result will also be stated in terms of the geometrodynamical
variables $(g_{ij}$, $\pi^{ij})$.

     In four spacetime dimensions, when the manifold is $R\times
\Sigma$ and the spatial slice $\Sigma $ is asymptotically flat at

infinity, the expression for total angular momentum is
well-known;
it is a two-dimensional integral over the surface at spatial
infinity \cite{RTei}.  In contrast, conserved quantities in
special
relativity  are typically three-dimensional integrals over the
volume of $\Sigma$.  The intuitive reason why conserved
quantities
in general relativity are  associated with surfaces, rather than
volumes, is that conserved quantities are associated with
coordinate transformations, via Noether's Theorem; and in general
relativity coordinate transformations within the volume have
little
physical meaning because of the diffeomorphism invariance of the
theory.   The transformations at the surface must preserve the
asymptotic flatness of the theory, and are reminiscent of
transformations in special relativity; consequently the
associated
conserved quantities depend only on degrees of freedom at the
surface.  In the planar case (effectively one-dimensional because
there is no x,y dependence) a ``volume'' integral is a one-
dimensional integral along z, from a left boundary $z_l$ to a
right
boundary $z_r$.  The ``surface'' is the two points $z_l$ and
$z_r$.
Surprisingly, the total spin momentum in the planar case has  the
form of a volume integral.

     In certain respects, however,  the planar case resembles
special
relativity more than general relativity.  In practice one fixes
the
variables x,y (also the variables X,Y in the local
Lorentz frame) so that the symmetry in the xy plane is manifest;
consequently tensors which have only x, y, X, and Y indices
(``transverse'' tensors) are left invariant by the surviving
diffeomorphisms and the local Lorentz transformations which have
not
been fixed.   The transverse tensors therefore resemble scalar
fields in special relativity, rather than tensors in general
relativity.  It turns out that the expression for spin angular
momentum contains only transverse tensors.

     The remainder of this section introduces the Ashtekar
notation
and writes out the expression for spin angular momentum in both
Ashtekar and geometrodynamical language.  Section \ref{der}
derives
this expression; section \ref{con} examines the connection
between
spin angular momentum and four conserved quantites previously
derived by Husain and Smolin \cite{HSm}.

     The basic variables of
the Ashtekar approach are an inverse densitized triad \E{a}{A}
and a complex SU(2) connection \A{A}{a}.
\begin {eqnarray}
     \E{a}{A}& =& e e^a_A; \\
\label{eq:1.1}
     [\E{a}{A},\A{B}{b}]&=& \hbar \delta (x-x') \delta ^B_A
\delta ^a_b.
\label{eq:1.2}
\end{eqnarray}
All quantities are three-dimensional unless explicitly indicated
otherwise.   Upper case indices A,
B, $\ldots $,I, J, K, $\ldots$ denote local Lorentz indices
("internal" SU(2) indices) ranging over X, Y, Z only.  Lower case
indices a, b, $\ldots $, i, j, $\ldots $ are also three-
dimensional and denote global coordinates on the three-manifold.
The quantity e is the determinant of the 3x3 spatial subblock of
the tetrad matrix $e_a^A$; similarly $e^a_A$ is an  inverse
tetrad.  I
use Levi-
Civita symbols of various dimensions: $\epsilon _{TXYZ} =
\epsilon _{XYZ} = \epsilon _{XY} = +1$.

     The planar symmetry (two spacelike commuting Killing
vectors, $\partial_x$ and $\partial_y$ in appropriate
coordinates)
allows Husain and Smolin \cite{HSm} to
solve and eliminate four  constraints (the
x and y vector constraint and the X and Y Gauss constraint) and
correspondingly eliminate four pairs of  (\E{a}{A}, \A{A}{a})
components.  The 3x3 \E{a}{A}\
matrix then assumes a block diagonal form, with one 1x1 subblock
occupied by \E{z}{Z}\, plus one 2x2 subblock which contains all
the ``transverse''  \E{a}{A},  those  with a = x,y and A =
X,Y.  The 3x3 matrix of connections
\A{A}{a}\ assumes a similar block diagonal form.  None of the
surviving fields depends on x or y.

     In the Ashtekar notation the total spin angular momentum
around the z axis is given by the quantity
\be
     L_Z =i \int dz [\E{y}{I} (\A{I}{x} - Re \A{I}{x})
                    - (x \leftrightarrow y)]
\label{eq:c9}
\ee
The integral is over the entire wave
packet, that is from $z_l $ to $z_r $.  I have chosen the
notation $L_Z $ rather than $L_z $ deliberately, since $L_z $
suggests a covariant vector, whereas \eq{c9} defines an
invariant.  The corresponding integral in geometrodynamical
notation is
\be
     L_Z = 2 \int dz [g_{xj}\pi^{yj} - g_{yj}\pi^{xj}],
\label{eq:1.4}
\ee
where $g_{ij} $ is the three-metric and $\pi^{ij}$ its conjugate
momentum.

   The integrals in \eqs{c9}{1.4} range over $z_l \leq z \leq
z_r$,
where the left and right boundary points $z_l$ and $z_r$ are a
{\it
finite} distance from the origin.  In three spatial dimensions it
is usual to place the
boundary surface at spatial infinity.  Bringing the surface at
infinity in to finite points is a major change, because at
infinity the metric goes over to flat space, and flat space is a
considerable simplification.  In the present case (effectively
one dimensional because of the planar symmetry)  the space does
{\it not} become flat at z goes to infinity, and nothing is lost
by
considering an arbitrary location for the boundary surface.
The result that the space
does not become flat as z goes to infinity was established in
paper
II.  Note that this result agrees
with one's  intuition from Newtonian gravity, where the potential
in one spatial dimension due to a bounded source does
not fall off, but grows as z at large z.

     The planar metrics considered here and in previous work
\cite
{I,II,III} admit two null
vectors k and l which have the
right hypersurface orthogonality properties to be the propagation
vectors for right-moving (k)
and left-moving (l) gravitational waves along the z axis.  It is
possible to choose z and t coordinates so that the equations
defining these hypersurfaces become especially simple: u = (ct -
z)/$\sqrt{2}$ = constant and v = (ct + z)/$\sqrt{2}$ = constant
\cite{Szmet}.  I have not used these special coordinates in the
proof, and one might ask whether the angular momentum exists for
more general metrics which are planar symmetric but do not admit
hypersurface orthogonal null congruences \cite{EhlK}.  I believe
the answer is yes, but have not checked whether the gauge fixing
employed by Husain and Smolin \cite{HSm} continues to go through
for the more general cases.

     I have also used the assumption that the gravitational wave
is
confined to a wave packet which lies  entirely inside the
boundary and
has not reached the boundary points $z_l$, $z_r$.  This
assumption
is used in a subtle way.  The derivation in section
\ref{der}
uses the Noether procedure, which in turn assumes that a
Lagrangian
formulation exists.  When the space is non-compact, as it is here
(the z axis is the entire real line, not a circle), then the
Lagrangian must contain surface terms, or the Euler-Lagrange
procedure is not well-defined \cite{II, RTei}.  In reference
\cite{II} I constructed the appropriate surface term, assuming
that
certain degrees of freedom are absent at the boundaries $z_l$,
$z_r$.  These degrees of freedom produce the transverse
displacements of test particles characteristic of gravitational
radiation.  In the language of reference \cite{II}, the fields B
and W were assumed to vanish at the boundaries.  (The assumption
that B and W vanish at
the boundaries  may be restated in a more covariant language: the
Weyl tensor has components which produce transverse deviations of
geodesics, and these components vanish at boundaries.)  I was
unable
to
construct a suitable surface term, therefore was unable to define
a Lagrangian, for B or W non-zero at the boundaries.

\section{Derivation}
\label{der}

     To prove \eq{c9}, I use the Noether procedure, with some
modifications \cite{RTei}.  In the usual, three-dimensional case,
one
exploits the invariance of the action under general coordinate
transformations $x^{\mu} \rta x^{\mu} + \xi ^{\mu} $, where $\xi
^{\mu} $ at infinity must reduce to a rotation around z:
\bea
          \xi ^b &=& \epsilon_{bzc} \delta \phi ^z x^c ; \nn
             b,c &=& x,y; \nn
     \delta \phi ^z &\rta & \mbox{constant}.
\label{eq:c10}
\eea
In the present, one-dimensional case, $\delta \phi ^z  $ must be
a rigid rotation, that is,
a constant for all z, or the transformation generated by $\xi $
will destroy the gauge conditions on the fields \A{A}{a}.  (The
matrix \A{A}{a} is block diagonal, with off-diagonal elements
\A{Z}{x} = \A{Z}{y} = \A{z}{X} = \A{z}{Y} = 0.)  The change in
\A{I}{i} is given by (minus one times) the Lie derivative.
\be
     \delta \A{I}{i} = -\xi ^b \partial_b \A{I}{i}
                              - (\partial_i \xi^b ) \A{I}{b}.
\label{eq:c11}
\ee
If $\xi $ depends on z, say, then the second term on the right
will allow $\delta \A{I}{z} $ to be non-zero for I = X,Y, which
violates the gauge conditions.  The first term on the right
cannot cancel the second, because the first term vanishes: b = x
or y, and the fields do not depend on x or y.  Thus $\xi
$ must be a constant for all z.  This result is consistent with
the observation made in section \ref{int}: the x,y sector of the
theory resembles special relativity rather than general
relativity.

     Now continue with the usual Noether procedure.  The
variation in the action is (suppressing the obvious indices for
simplicity)
\be
     \delta S = \int dt dz [i\Etld \delta \dot{A}
               + i \delta \Etld \dot{A}
               - (\delta H/ \delta \Etld) \delta \Etld
               - (\delta H/ \delta A) \delta A ].
\label{eq:c12}
\ee
Because of the classical equations of motion, the second term in
the square brackets cancels the third, while the fourth term
provides an $i \partial_t \Etld \delta A $ term which combines
with the first term.  Also, because the action is invariant under
the transformation generated by \eq{c10}, $\delta S = 0 $.
\Eq{c12} becomes
\bea
     0 & = & \int dt dz \partial_t [ i \Etld \delta A ] \nn
      & = & \int dt dz \partial_t [i \E{a}{I} ( - \A{I}{b}
                    \partial_a \xi^b ) ]     \nn
     & = & \int dz [i \E{a}{I} (- \A{I}{b} \epsilon_{bza}
               ]_{t1}^{t2} \delta \phi ^z \nn
     & =: & [L_Z (t_2) - L_Z (t_1) ] \delta \phi ^z .
\label{eq:c13}
\eea
On the second and third lines I have used \eqs{c10}{c11}, with i
=
x,y,z replaced by a = x,y only.

     The quantity $ L_Z $ on the final line is not quite the
desired constant of the motion, \eq{c9}: since the original
action S was complex, the $ L_Z $ in \eq{c13} is also complex.  I
wish to argue that I can and should drop the imaginary part of $
L_Z $.  Normally the complex action is assumed to depend on both
the connection A and the densitized triad \Etld, which are varied
as independent fields.  Before the 3+1 reduction, the complex
action depends on the four-dimensional spin connection $
\omega^{IJ}_i $ and the tetrad $ e^I_i $, also varied as
independent fields.  So long as $ \omega $ and e are treated as
independent fields, the imaginary part of S is non-trivial.  If
one invokes the classical equations of motion, however, these
give the usual relation $ \omega = \omega (e) $ between tetrad
and connection.  Once this relation is inserted back into the
action, Im S vanishes trivially because of the Bianchi identity.
In the Noether procedure I assume the classical equations of
motion.  Therefore Im S vanishes, and I should discard the
imaginary part of the constant of motion in \eq{c13}.  This
leaves me with the constant of the motion given at \eq{c9}.
$\Box$

     I should perhaps point out that in the usual, three-
dimensional derivations of conserved quantities, the $ \partial_i
$ and $ \partial_b $ in \eq{c11} are manipulated until the
integrand in \eq{c13} becomes a total derivative $ d^3x
\partial_i [\cdots] $.  Integration by parts with respect to $
dx_i $ then leads to a constant of the motion in the form of a
surface term.  This procedure is not possible here because i =
x,y only, and there are no dx or dy integrations.

     I have tacitly dropped an ``orbital'' contribution to $ L_Z
$, so that
this quantity is pure spin angular momentum.  At \eq{c11} I
dropped the $ \xi ^b
\partial_b $ term because b = x,y only.  From \eq{c10} $
\xi ^b \partial_b $ is essentially $ (r \times \bigtriangledown
)_z $, the orbital angular momentum.

     The expression \eq{c9} can be quantized readily.  In
references
\cite{II,III} I constructed solutions $\psi$ which were
annihilated
by all the constraints and which depended only on the transverse
fields \E{a}{A} and the connection \A{Z}{z}.  In order to
represent the commutation relations
\eq{1.2} correctly, one can make the remaining
fields into functional derivatives in the standard manner,
\bea
     \E{z}{Z} &\rta& -\hbar \delta / \delta \A{Z}{z}; \nn
     \A{A}{a} &\rta& +\hbar \delta / \delta \E{a}{A}, \nn
               & &  \mbox{(for a = x,y and A = X,Y)},
\label{eq:2.0}
\eea
then order functional derivatives to the right in \eq{c9}.

     From the discussion so far, it is not obvious that $L_Z$ is
the intrinsic spin operator for a
helicity two field.  The pattern of x and y indices in \eq{c9}
looks like a spin one cross product.
To clarify the helicity content, it is helpful to express $L_Z$
in terms of fields which are
eigenstates of rotations around the Z and z axes.   In  three
dimensions, the local Lorentz
symmetry is O(3) (rather than
SU(2), because there are only vector, not spinor indices).
After  the X and Y
internal Gauss constraints are fixed, O(3) reduces to O(2), the
group of rotations about Z.   It is
convenient
to shift to transverse fields which are eigenstates under these
rotations:
\begin{equation}
     \E{a}{\pm} = [\E{a}{X} \pm i\E{a}{Y} ]/\sqrt{2} ,
\label{eq:1.3}
\end{equation}
where a = x,y; and similarly for \A{\pm}{a}.  Expand
not only the local indices X, Y, but also the global indices
x, y, because  after gauge fixing the latter
indices possess a residual O(2) symmetry.  This symmetry, a
rigid rotation around the z axis which makes no
distinction between contravariant and covariant indices x and y,
is just the symmetry which was
used originally to construct $L_Z$ via the Noether procedure.
When the fields in \eq{c9} are
expanded in O(2) eigenstates, one gets (after some manipulations;
see the next paragraph)
\bea
     L_Z &=& 2\int dz [\E{+}{+} \A{-}{-} - \E{-}{-} \A{+}{+}] \nn
          &=& 2 \hbar \int dz [\E{+}{+} \delta / \delta \E{+}{+}
                     - \E{-}{-}  \delta / \delta \E{-}{-}].
\label{eq:c26}
\eea
This expression counts the number of \E{\pm}{\pm} fields in the
wavefunctional $\psi$,
assigning each field a value $\pm2\hbar$.  This looks very much
helicity two, and implies that
the helicity content of $\psi$ is
determined by the number of transverse $\E{\pm}{\pm}$ fields that
it contains, but not by its
$\E{\pm}{\mp}$ fields.  Similarly, if one  uses
a connection representation
rather than a momentum representation, the helicity content is
determined by the number of
transverse $\A{\pm}{\pm}$ fields.

     Proof of \eq{c26}: I wish to manipulate the Re A term in
\eq{c9} extensively, while
leaving alone the first two terms.  Accordingly I introduce the
following notation which allows
me to abbreviate the first two terms.
\be
     G^a_b = \int \E{a}{I} \A{I}{b}.
\label{eq:c13'}
\ee
Then
\be
     L_Z = -i \epsilon_{ab}G^a_b +i \int dz \epsilon_{ab}\E{a}{I}
          {\mbox Re} \A{I}{b}.
\label{eq:c21}
\ee
Before this expression can be quantized, the Re A term must be
written out in terms of  \Etld\
fields.  The following formulas are useful.
\bea
      2 ^{(4)}A^{MN}_b &=& \omega ^{MN}_b  + i
\epsilon^{MN}_{..PQ}
                                    \omega ^{PQ}_b /2 ; \nn
     \A{I}{b} &=&  ^{(4)}A^{MN}_b \epsilon_{MIN} ; \nn
           Re  \A{I}{b}&=&\epsilon _{IJ} \omega^{ZJ}_b ;\nn
     \omega ^{IJ}_b &=& [\Omega _{i[jb]} + \Omega _{j[bi]}
                         - \Omega_{b[ij]}]e^{iI} e^{jJ};\nn
     \Omega _{i[jb]} &=& e_{iM}[\partial _je^M_b
                                   - \partial_be^M_j]/2.
\label{eq:2.19}
\eea
The first two lines relate the four dimensional complex
connection to the corresponding three
dimensional connection, and to the four dimensional Lorentz
connection $\omega ^{MN}_b$.  I
use the same sign conventions as in reference \cite{II}.  The
third line  is the
reality condition for the transverse
connections  \A{I}{b}.    $e_{iM}$ is the
tetrad, and $
e^M_i $ is its inverse.   The standard \cite{Pel} Lorentz gauge
fixing
condition, $e^{tZ} = 0$, may be used
to simplify the sums over i in the definition of $\omega
^{IJ}_b$,
when I = Z; one gets
\bea
      \omega ^{ZJ}_b &=& -\partial _zg_{bj}e^{zZ}e^{jJ}/2; \nn
     \E{a}{I} Re \A{I}{b} &=& e e^a_B \epsilon _{BJ}
                    \omega^{ZJ}_b \nn
                    &=& -\partial _zg_{bj} \epsilon^{aj}/2,
\label{eq:2.20}
\eea
which turns out to be  a total derivative.  Now convert back to
the basic Ashtekar fields, the
densitized triads.  A useful  identity is
$g_{bj} = \epsilon_{bm} \epsilon_{jn} g^{mn} g^{(2)}$, where  all
metric components and
determinants are in the 2x2 transverse sector.  One needs also
the definition of the
densitized triads, \eq{1.1}.  The Re A term in \eq{c21} becomes
\be
     \E{a}{I}{\mbox Re} \A{I}{b} = - \partial_z [\E{m}{M}
\E{a}{M}
               \E{z}{Z} /\Etwo] \epsilon_{bm} /2,
\label{eq:c22}
\ee
\Etwo  is the determinant of the matrix \E{a}{I} in the 2x2
transverse sector.  When this
expression is inserted into \eq{c21}, the Re A  term becomes
\be
     -i \int \partial_z [\E{m}{M} \E{m}{M} \E{z}{Z} /\Etwo] /2.
\label{eq:c23}
\ee
This expression is  quantized by replacing the \E{z}{Z} by a
functional
derivative, as at \eq{2.0}.  The
remaining functions, the transverse \Etld, remain functions.
Since the expression \eq{c23} is a
total derivative, one may replace these functions by any
expression which has the same limit at z
= $z_l$ or $z_r$.  At the boundaries, the transverse \Etld\
become
conformally flat \cite{II}: \E{m}{M} $\rta$
(conformal factor) $\times \delta ^m_M$.  Therefore one may
replace
\be
     \E{m}{M} \E{m}{M}  /\Etwo \rta 2.
\label{eq:c24}
\ee
\Eq{c23} becomes
\bea
    i \int dz \epsilon_{ab} \E{a}{I}{\mbox Re} \A{I}{b}& =&  -i
\int \partial_z \E{z}{Z} dz
                                                  \nn
               &=& -i \int \epsilon_{MN} \E{c}{M} \A{N}{c},
\label{eq:c25}
\eea
where I have used the surviving Gauss constraint,
$\partial_z\E{z}{Z} -\epsilon_{MN}
\E{a}{M}
\A{N}{a} = 0$, to replace the
\E{z}{Z} by transverse fields.
Now insert this back into $L_Z$, \eq{c21}.  At this point the
expression no longer involves Re
A.   Expand in terms of O(2) eigenstates, remembering to contract
every + index with a - index,
as $\epsilon_{MN} \E{a}{M} \A{N}{a} = \epsilon_{+-} \E{a}{-}
\A{+}{a} + \cdots $, with
\be
     \epsilon_{\pm \mp} = \mp i.
\label{eq:c25'}
\ee
The result is  \eq{c26} $\Box$

       In \eq{c9} or \eq{c21}, the
term containing  Re A is separately a constant of the motion.
Proof: the term is obviously a
scalar under internal Gauss rotations about Z.  When the term is
commuted with the remaining,
scalar and z diffeomorphism constraints, those constraints must
be smeared with a functions
which represent small changes in the lapse and shift.  The
smearing functions must vanish at
boundary points
\cite{II, III}, since lapse and shift are required to reduce to
fixed constants there, and cannot
change at boundaries.   Hence the scalar and diffeomorphism
constraints  commute with any
expression which depeds only on  fields evaluated at boundary
points.  But  the integrand of the
Re A term is a total derivative, \eq{c23}, therefore the term
depends only
on fields evaluated at the boundaries, z = $z_l$ and $z_r$.
In fact any surface term which is
Gauss invariant is automatically a
constant of  the motion. $\Box$

     If the Re A term in \eq{c21} is separately a constant of the
motion, then the quantity
$\epsilon_{ab}G^a_b$ in that equation is also separately a
constant of the motion, where
$G^a_b$ is the integral defined at \eq{c13'}.  Husain and Smolin
have shown that all four
$G^a_b$ integrals are constants of the motion \cite{HSm}.  These
authors apparently did not use
Noether's theorem, so were
unaware of the connection with total spin.  It is possible to
recover two more of the Husain-Smolin conserved quantities by
using a Noether procedure, since the Lagrangian happens to be
form-invariant under the larger group SL(2,R).  (The group acts
on the (x,y) indices, like the O(2) group used  in the derivation
of $L_Z$ .  Covariant indices are transformed by an SL(2,R)
matrix,
while contravariant indices are transformed by another matrix
which is the transposed inverse of the matrix for the covariant
indices.)

     I have tried to derive the fourth Husain-Smolin
conserved quantity by enlarging SL(2,R) to GL(2,R), i.\ e.\ by
adding dilatations.  However, there is a technical difficulty.
Since the Lagrangian is a density rather than a scalar, it is not
form-invariant under dilatations.  Normally the Lagrangian is
multiplied by a dxdy integration which is also non-invariant, so
that the action as a whole is invariant.  In the present case,
however, there is no dxdy integration, and it is not clear how to
apply the Noether procedure.  The Husain-Smolin quantities are
discussed further in the next section.

\section{Discussion}
\label{con}

     Husain and Smolin have shown that all four
of the integrals  $G^a_b$, \eq{c13'}, are conserved \cite{HSm}.
The present work gives a
physical interpretation to one linear combination of these
integrals, $\epsilon_{ab}G^a_b =
G^x_y - G^y_x$ (more precisely, to the imaginary part of this
linear combination).  From \eq{c21}, $L_Z$ equals
\be
     L_Z = -i [G^x_y - G^y_x - Re (G^x_y - G^y_x)].
\label{eq:6.1}
\ee
It is possible to find physical interpretations for two other
linear combinations of the $G^a_b$.   From section \ref{der},
after quantization \eq{6.1} can be rewritten as at \eq{c26}:
\be
     L_Z = 2 \hbar \int dz [\E{+}{+} \delta / \delta \E{+}{+}
                     - \E{-}{-}  \delta / \delta \E{-}{-}].
\label{eq:6.2}
\ee
The $\pm$ indices refer to the O(2) eigenstates defined at
\eq{1.3}.  As remarked in section
\ref{der}, this expression implies that the spin content of any
wavefunctional $\psi$ is
determined by the number of transverse $\E{\pm}{\pm}$ fields that
$\psi$ contains; $\E{\pm}{\mp}$ fields do not contribute to the
total spin.  Now express two
other linear combinations of $G^a_b$ components in terms of O(2)
eigenstates:
\bea
     G^x_x - G^y_y \pm i (G^x_y + G^y_x) &=&
                              2\int dz \E{\pm}{I} \A{I}{\mp} \nn
          &=& 2\hbar \int dz \E{\pm}{I} \delta / \delta
\E{\pm}{I}.
\label{eq:6.3}
\eea
These operators are raising and lowering operators for intrinsic
spin.  The upper sign (for example) replaces
\bea
     \E{-}{+} \rta  2 \hbar \E{+}{+}; \nn
     \E{-}{-} \rta  2 \hbar \E{+}{-},
\label{eq:6.4}
\eea
when acting on a solution  $\psi$, hence raises the $L_Z$
eigenvalue by $2
\hbar $ units.

     The remaining linear combination, $ G^x_x + G^y_y $, is a
two
dimensional version of a three-dimensional operator which plays a
key role in Thiemann's regularization scheme \cite{Thie}.  It is
also a number
operator for the number of transverse \Etld  fields in $\psi$: if
$\psi$ is a string of n \Etld\  fields, then the eigenvalue of $
G^x_x + G^y_y $ is $\hbar n$.  Presumably n characterizes  the
background geometry.  To clarify this, one can write out the
Ashtekar connections in $G^x_x + G^y_y $ in
terms of the four-dimensional Lorentz connection $\omega
^{IJ}_a$;
then use the classical equations of motion to express $\omega$ in
terms of the tetrads.  Use a conformally flat gauge to shorten
the
(lengthy but straightforward) algebra.  The necessary formulas
relating \A{A}{a} to $\omega$, and $\omega$ to the tetrads, are
given at \eq{2.19}.  The final
result is
\be
     G^x_x + G^y_y = i\int dz  \partial [e^{(2)}]/\partial t.
\label{eq:6.5}
\ee
In words, the average rate of change of the area operator is a
constant of the motion.

     As already stressed in section \ref{der},  if one
uses a connection
rather than a momentum representation, then the transverse
\A{I}{a}
fields play the role formerly played by the transverse \E{a}{I}.
In particular, the spin is determined by the \A{\pm}{\pm} fields;
and $ G^x_x + G^y_y $ counts the number of transverse A fields.

\end{document}